\documentclass[lettersize,journal]{IEEEtran}
\usepackage{amsmath,amsfonts}
\usepackage{algorithmic}
\usepackage{algorithm}
\usepackage{array}
\usepackage[caption=false,font=normalsize,labelfont=sf,textfont=sf]{subfig}
\usepackage{textcomp}
\usepackage{stfloats}
\usepackage{url}
\usepackage{verbatim}
\usepackage{graphicx}
\usepackage{cite}
\usepackage{amssymb}
\begin{document}

\title{FOSSA: First-Order Optimality–Based Sensor Selection for PINN Inverse Problems$,$ with Application to Electrocardiographic Imaging}

\author{Jianxin Xie$^1$
\thanks{$^{1}$University of Virginia, School of Data Science, email address: hcf7fd@virginia.edu}
}



\maketitle

\begin{abstract}

Physics-informed neural networks (PINNs) have emerged as a powerful framework for modeling physical systems and solving inverse problems. In such settings, sensors are deployed to capture observable system responses; however, the quality of reconstruction critically depends on how these sensors are selected. Existing sensor selection strategies for PINNs are closely related to active learning and experimental design, typically relying on iterative refinement schemes that sequentially add sensors and retrain the model. While effective under limited data regimes, these approaches incur substantial computational cost due to repeated retraining and primarily focus on selecting subsets of sensors, without providing a global characterization of sensor importance.
In this work, we propose FOSSA, a first-order optimality-based sensor selection algorithm for inverse PINNs. Unlike existing methods, FOSSA evaluates sensor importance in a post-training manner, requiring only a single trained PINN. FOSSA assigns importance scores to all candidate sensing locations based on the first-order optimality condition at convergence. To improve robustness, a refinement scheme is further proposed to handle instability in the inverse solver. FOSSA facilitates a global assessment of the contribution of each sensor to reconstruction. We validate the proposed approach on the inverse electrocardiography (ECG) modeling and show that not all sensors contribute positively to predictive performance. Incorporating low-importance sensors can, in fact, degrade reconstruction accuracy. These findings highlight the need for principled sensor importance evaluation and provide a scalable pathway for guiding sensor deployment in physics-informed inverse modeling.
\end{abstract}


\section{Introduction}
Inverse modeling is ubiquitous in scientific and engineering applications, where the objective is to recover latent system states from indirect observations \cite{pullan2010inverse, baillet2001evaluation, maniatty1989finite}. In practice, sensor observations of such systems are often indirect, noisy and limited due to constraints in measurement accessibility, sensor availability, or cost, leading to ill-posedness where the predictive solution becomes unstable and non-unique within a large feasible space \cite{peng2021role, tamarozzi2016noise}. As a result, the accuracy of the inferred solution is fundamentally constrained not only by the amount of data, but also by the informativeness and placement of the sensors. 
With partial coverage measurements, purely data-driven approaches are prone to overfitting noise and may yield physically inconsistent solutions, underscoring the fundamental limitations imposed by sparse sensing in inverse problems \cite{arridge2019solving}.

Physics-informed neural networks (PINNs) provide a principled framework to address the challenge by incorporating governing physical laws directly into the learning process \cite{raissi2019physics,raissi2024physics,karniadakis2021physics}. Instead of relying solely on observational data, PINNs enforce the prediction with constraints derived from well-established physics governing rules. This integration effectively regularizes the solution space by eliminating physically infeasible candidates, thereby improving robustness against noise and data sparsity. As a result, PINNs have demonstrated strong potential in solving inverse problems across various domains, such as fluid dynamics \cite{sallam2023use, cai2021physics1}, heat transfer \cite{cai2021physics2, oommen2022solving}, geophysical inversion \cite{rasht2022physics}, and biomedical imaging \cite{banerjee2024pinns}, where the combination of limited observations and known physical mechanisms can be leveraged to achieve more accurate and stable reconstructions.

Despite the promising performance of PINNs in inverse problems, most existing studies implicitly assume that sensor observations are given and fixed, without considering how the measurements themselves are acquired \cite{xie2022physics, oommen2022solving, chiu2022can}. In practical settings, however, the number and placement of sensors are often strictly constrained, and the quality of the inverse solution depends critically on the informativeness of these measurements. As such, identifying regions that exhibit high information flow from the observation domain (i.e., the sensor-accessible outer surface) is essential for optimizing reconstruction outcomes. Conventional sensor placement strategies, such as uniform sampling or geometry-based heuristics 
\cite{zabini2016random, johnson1990minimax}, do not account for the underlying physical dynamics or the learning process of the model. As a result, these approaches may select sensors that are suboptimal for reconstructing the latent physical states. 

Existing efforts on sensor selection in PINNs are closely related to active learning and experimental design strategies for inverse problems. Many of these approaches adopt an iterative refinement paradigm, where a PINN model is first trained using an initial set of observations, and additional sensing locations are selected based on criteria such as residual magnitude, predictive uncertainty, or information gain \cite{tristani2025active, gao2023active, xie2022physics1, wang2024optimal}. The selected sensors are then incorporated into the training set, and the model is retrained to update the solution, forming a sequential loop of sensor acquisition and model refinement. While effective under limited data regimes, such strategies often incur substantial computational cost due to repeated retraining. 
Moreover, these methods focus on selecting subsets of sensors, but do not provide a global characterization of sensor importance across the domain, particularly in settings where a dense set of candidate locations is available and the goal is to understand their relative contributions. Nor do they explicitly leverage the training convergence of the underlying inverse model. As a result, they offer limited insight into how different regions contribute to reconstruction, which is critical for guiding principled sensor deployment and identifying intrinsically informative areas.

To address these limitations, we propose a generic framework for sensor selection based on importance scoring in PINN-based inverse problems. Built upon the first-order optimality condition, the proposed method evaluates sensor importance through the optimization dynamics of the trained model. In contrast to existing approaches that rely on iterative retraining, our framework operates in a post-training manner and requires only a single trained PINN model, enabling efficient and scalable sensor selection. We exemplify the proposed framework using the inverse electrocardiographic imaging (ECGI) problem, where hundreds of candidate sensors are distributed on the body surface, providing a rich setting where the wise sensor selection can substantially improve the reconstruction accuracy of the heart dynamics. The main contributions of this work are summarized as follows:
\begin{itemize}
\item A general post-training framework for sensor selection in PINN-based inverse problems, eliminating the need for iterative retraining and enabling efficient evaluation of sensor importance.
\item A first-order optimality-driven importance metric, which quantifies the global contribution of each sensor through the coupled data and physics constraints.
\item A robust importance estimation strategy with confidence-aware refinement, improving stability and reliability in practical settings with noisy and limited observations.
\end{itemize}

Our results demonstrate that, under moderate noise levels, incorporating low-quality sensors (i.e., those with low importance scores) can in fact degrade reconstruction accuracy.


\section{Literature Review}
\label{Section:Lit}

Complex physical systems are often governed by high-dimensional and nonlinear dynamics, making accurate reconstruction inherently challenging. In practice, such systems are only partially observed through a limited number of sensors, which are often subject to measurement noise and spatial sparsity. Under these constraints, recovering the full system state from incomplete and noisy observations remains highly ill-posed, leading to significant degradation in reconstruction accuracy \cite{santos2023development, collins2023super, karniadakis2021physics}. Recently, PINNs have been proposed as a powerful framework for modeling complex physical systems by embedding governing physics laws directly into the learning process, typically in the form of partial differential equations (PDEs) \cite{raissi2019physics, karniadakis2021physics, farea2024understanding}. In this setting, the solution is not solely driven by sensor observations, but is also constrained to satisfy the underlying physical principles. Such integration substantially alleviates the dependence on dense sensor measurements while improving reconstruction fidelity through physics-consistent regularization \cite{karniadakis2021physics, ren2023physr}. PINNs have been successfully applied across a wide range of domains, including fluid dynamics \cite{cai2021physics, raissi2020hidden, ali2025machine}, bioelectrical systems \cite{zhu2025physics, xie2022physics, xie2022physics1, zhao2025physics}, and material science \cite{haghighat2021physics, kim2025physics, nguyen2024parcv2}, among others.

In many practical applications, PINNs are employed to solve inverse problems, where the goal is to infer hidden system states from partial observations \cite{raissi2024physics, yuan2022pinn, depina2022application, baldan2023physics}. A representative example is the inverse electrocardiography (ECG) problem, which aims to reconstruct cardiac electrical dynamics on the heart surface from multi-channel body surface potential measurements \cite{xie2022physics, zhu2025physics, herrero2022ep }. In this setting, the quality and configuration of sensors play a critical role in determining reconstruction accuracy. Despite the physics constraints imposed by PINNs, insufficient or poorly distributed sensors can still lead to large uncertainty and degraded solutions, particularly for highly ill-posed systems \cite{toloubidokhti2025meta, li2024solving, xie2022physics, zhu2025physics}. Therefore, beyond model design, the selection and placement of sensors become a key factor in improving inverse problem performance. However, most existing sensor selection frameworks assume a fixed sensor configuration and do not explicitly address how sensor locations or subsets should be optimized to maximize reconstruction fidelity \cite{santos2023development,liu2024physics,jekic2025examining}.


The problem of sensor selection is fundamental in scientific computing and data acquisition, and has been extensively studied in experimental design and related fields \cite{huan2024optimal, manohar2018data}. Common strategies include uniform sampling, where sensors are evenly distributed across the domain \cite{santner2003design, bash2004approximately}, and random sampling, which serves as a simple baseline without structural assumptions \cite{zabini2016random, bopardikar2017sensor}. Geometry-based heuristics have also been widely adopted, such as maximizing spatial coverage or enforcing a maximin distance criterion to promote diversity among sensor locations \cite{johnson1990minimax, pronzato2017minimax}. In addition, clustering-based methods have been proposed to select representative sensors from spatial partitions \cite{menneer2023cluster, wohwe2019optimized, jia2015dynamic}, while information-theoretic approaches aim to maximize criteria such as entropy or mutual information \cite{ertin2003maximum, yan2020sensor, fonollosa2013temperature}. While these methods are effective in promoting spatial diversity or statistical representativeness, they are largely agnostic to the underlying physical dynamics of the system. More importantly, they do not account for the interaction between sensor observations and the model learning process, particularly in physics-informed settings where reconstruction is jointly governed by data fidelity and physical constraints.

For physics-governed systems, sensor selection is influenced not only by the geometric configuration of the domain, but also by the underlying physical dynamics and the model updating process during inverse problem solving. Existing approaches in the PINN literature are closely related to adaptive sampling, where informative sensor locations are typically identified through iterative refinement. For example, Lu et al. introduced residual-based adaptive refinement (RAR) within the DeepXDE framework, which iteratively adds new collocation points at locations where the PDE residual is largest, progressively concentrating computational effort in under-resolved regions of the domain \cite{lu2021deepxde}. Wu et al. further extended this idea into residual-based adaptive distribution (RAD and RAR-D), which instead resample all training points from a probability density function proportional to the PDE residual, providing a more globally balanced refinement strategy \cite{wu2023comprehensive}. However, these methods focus on refining collocation points for improving PDE solution accuracy, rather than selecting or optimizing observation locations in inverse problems.

Beyond residual-driven selection, uncertainty also serves as an important indicator for guiding sensor placement. Regions with high predictive variance indicate insufficient observational constraints, and placing sensors in these areas can effectively improve the identifiability of the inverse solution. 
 Tristani et al. proposed a PINN framework integrated with active learning via Monte Carlo dropout to quantify epistemic uncertainty sequentially, querying new sensor measurements in regions where the model is least confident and iteratively refining both the sensor set and the network parameters \cite{tristani2025active}. Similarly, Gao \cite{gao2023active} proposed an active learning scheme for high-dimensional PDEs where new sample points are selected via a Gaussian process surrogate trained on the current PINN residual field, effectively coupling the surrogate model's uncertainty with physics-constraint violations to guide sample acquisition. 
 Xie et al. \cite{xie2022physics1} proposed a generic iterative algorithm to find the next best sensor locations for reconstructing whole-heart electrical dynamics, building a novel metric that jointly considers uncertainty distribution over the domain and a maximin geometric diversity criterion, realizing faster error reduction. Information-theoretic objectives have also been widely explored for sensor placement. For example, Wang et al. \cite{wang2024optimal} propose an optimal sensor placement algorithm based on mutual information and correlation, where sensor locations are selected by maximizing the mutual information between the chosen sensors and the uninstrumented candidate locations.



Recent work has explored formulating sensor selection as a global optimization problem within the PINN framework. For example, PIED \cite{hemachandra2025pied} proposes a differentiable experimental design strategy that optimizes sensor locations by minimizing the expected inverse error using PINNs as both forward simulators and inverse solvers. It further incorporates meta-learning to accelerate training and employs approximations to reduce the cost of evaluating candidate sensor configurations. However, the method still requires repeated training or approximation of multiple PINN instances across different parameter realizations, leading to considerable computational overhead, and operates in a pre-training setting without explicitly leveraging the sensitivity structure of a trained model.

Despite these advances, existing approaches primarily focus on directly selecting a subset of sensors based on predefined criteria, without providing a global characterization of how different sensing locations contribute to the inverse reconstruction. In many practical scenarios, especially in early-stage system design or prototyping, it is often feasible to assume access to a dense or full set of candidate sensing locations. Under such a setting, the goal shifts from selecting a minimal subset to understanding the relative importance of sensors across the entire domain, in order to guide subsequent deployment strategies. This perspective enables the identification of regions that are inherently more informative for reconstructing the underlying physical state, which can later inform decisions such as allocating higher sensor density in critical areas. Therefore, rather than performing direct sensor selection, it is equally desirable to investigate can evaluate sensor importance from the perspective of the model in the inverse problem structure, providing a principled foundation for downstream sensor deployment and system design.

\section{Methodology}
\label{Section: Method}

\subsection{Physics-Informed Neural Networks for Inverse Problems}
PINNs \cite{raissi2019physics,raissi2020hidden} have recently emerged as a powerful framework for solving scientific computing problems governed by physical laws. Unlike conventional neural networks that rely purely on data, PINNs incorporate known physical principles directly into the training process by embedding governing equations into the learning objective. This approach allows neural networks to learn physically consistent solutions even when only limited observational data are available.

PINNs have been widely used for both forward and inverse modeling. In forward problems, the governing equations and boundary conditions are known, and the objective is to compute the physical state of the system. In contrast, inverse modeling aim to infer unknown physical states from indirect and often sparse measurements. In many real-world systems, direct observation of the latent physical field is not possible, and measurements can only be obtained through a limited number of sensors that observe the system indirectly \cite{kapoor2023physics, xie2022physics, baillet2001evaluation}. Recovering the underlying state of the system from such measurements is typically ill-posed and highly sensitive to the quality and placement of sensors.

Consider a physical system governed by a differential operator:
\begin{equation}
    \mathcal{N}[u(x,t)]=0, \quad (x,t)\in \Omega \times [0,T]
\end{equation}

where $u(x,t)$ represents the unknown physical field defined over spatial domain $\Omega$ and time interval $[0,T]$, and $\mathcal{N}[\cdot]$ denotes the governing physics operator. This operator may include spatial derivatives, temporal dynamics, or nonlinear interactions depending on the underlying physical model. In inverse problems, the internal field $u(s,t)$ cannot be directly observed. Instead, measurements are obtained through a set of sensors that capture indirect observations of the system. Each sensor records measurements through an observation operator that maps the latent field to observable quantities. Formally, the measurement obtained from sensor $s$ can be written as

\begin{equation}
\label{forward}
    y_s=\mathcal{F}_s(u)+\epsilon_s,\quad s\in S
\end{equation}
where $y_s$ denotes the measurement recorded by sensor $s$, $\mathcal{F}_s(\cdot)$ represents the observation operator associated with that sensor, $S$ denotes the set of available sensors, and $\epsilon_s$ represents measurement noise. Within the PINN framework, the unknown physical fields $u(x,t)$ is approximated using a neural network $u_{\boldsymbol{\theta}} (x,t)$, where $\boldsymbol{\theta}$ denotes the trainable parameters of the network. The network takes spatial and temporal coordinates as inputs and outputs an approximation of the physical state at those locations. The parameters of the network are learned by minimizing a loss function that enforces both consistency with observational data and adherence to the governing physical laws.

Specifically, the training objective of the PINN is defined as a weighted combination of a data fidelity term and a physics constraint term
\begin{equation}
\mathcal{L}(\theta) = \lambda_d \mathcal{L}_{data} + \lambda_p \mathcal{L}_{phy}
\end{equation}
The physics term enforces the governing equations by penalizing violations of the physical law across the domain:
\begin{equation}
\mathcal{L}_{phy} = \|\mathcal{N}[u_{\boldsymbol{\theta}} (x,t)]\|^2
\end{equation}
The PINN will be trained until convergence, so that
\begin{equation}
\boldsymbol{\theta}^* = \arg\min_{\boldsymbol{\theta}}  \mathcal{L}(\boldsymbol(\theta))
\end{equation}

Through this optimization process, the neural network learns a representation of the physical field that both fits the available indirect measurements and satisfies the governing equations.

A critical challenge in PINN-based inverse problems lies in the selection and placement of sensors. Because measurements are often costly or limited in practice, only a subset of possible sensor locations can be used. The information provided by different sensors can vary significantly depending on how strongly their observations constrain the underlying physical field. Poorly chosen sensors may provide redundant or weakly informative measurements, leading to unstable or inaccurate reconstructions.

Therefore, determining which sensors provide the most informative observations is a key problem in inverse modeling. An effective sensor selection strategy should identify measurements that contribute the most to recovering the internal physical state under the PINN training objective. In the following section, we derive a principled sensor importance measure directly from the first-order optimality condition of the PINN, which forms the basis of the proposed FOSSA framework.

\subsection{FOSSA-Stage1: First-Order Optimality-based Sensor Importance Scoring}
We introduce FOSSA (First-Order Optimality-based Sensor Selection Algorithm), a principled framework for identifying informative sensors in physics-informed inverse problems. The key idea of FOSSA is to quantify the contribution of each observation to the final reconstruction accuracy by analyzing the optimality condition of the PINN training objective. Most existing sensor placement methods for inverse problems rely on geometric heuristics, greedy subset selection, or information-theoretic objectives such as maximizing mutual information between sensors and latent states \cite{tristani2025active, gao2023active, xie2022physics1, wang2024optimal}. While these approaches can improve coverage or statistical informativeness, they are largely decoupled from the optimization process of the inverse solver itself. In PINN frameworks, however, the reconstructed solution is obtained by minimizing a nonlinear objective that combines data mismatch and physical constraints. Consequently, the true contribution of each sensor depends on how its observation influences the optimization dynamics of the PINN model, a factor that is not captured by existing sensor placement strategies.

To formalize this idea, consider a PINN training objective composed of sensor-specific data mismatch terms and a physics constraint term. Let $N_b$ denote the number of candidate sensors available for observation. The weighted loss function can be written as
\begin{equation}
\label{weightedloss}
\mathcal{L}(\boldsymbol{\theta;w}) = \sum_{i=0}^{N_b}w_i l_i (\boldsymbol{\theta}) + \lambda  \mathcal{L}_{phy}(\boldsymbol{\theta})
\end{equation}

where $l_i(\boldsymbol{\theta})$ represents the data loss associated with sensor $i$, $w_i$ is a non-negative weight assigned to that sensor, and $\mathcal{L}_{phy}(\boldsymbol{\theta})$ denotes the physics-informed constraint term enforcing the governing equations. The scalar $\lambda$ controls the relative strength of the physics constraint. The individual data loss term is defined as
\begin{equation}
    l_i(\boldsymbol{\theta}) = \|y_i-\hat{y_i}(\boldsymbol{\theta})\|^2
\end{equation}
where $y_i$ is the observed measurement at sensor $i$. $\hat{y_i}(\boldsymbol{\theta})$ denotes the predicted measurement at sensor $i$, obtained by applying the forward transformation operator $\mathcal{F}_i$ to the estimated internal profile $\hat{u}_i$ produced by the neural network, i.e., $\hat{y_i}(\boldsymbol{\theta}) = \mathcal{F}_i(\hat{u}(\boldsymbol{\theta}))$. Introducing sensor weights allows us to analyze how the solution of the inverse problem changes when the contribution of each sensor is perturbed. 


Let $\boldsymbol{\theta}^*(\boldsymbol{w})$ denote the optimal network parameters obtained by minimizing this weighted objective. Because the training objective depends on the sensor weights $\boldsymbol{w}$, the optimal parameters implicitly become a function of $\boldsymbol{w}$. At convergence, the trained network satisfies the first-order optimality condition

\begin{equation}
\nabla_{\boldsymbol{\theta}} \mathcal{L}(\boldsymbol{\theta}^*(\boldsymbol{w}),\boldsymbol{w}) = 0.
\end{equation}

In the proposed framework, this optimality condition is used to perform a post-training sensitivity analysis that quantifies how perturbations in the weight of each sensor affect the learned model parameters and, consequently, the reconstruction error. To understand how each sensor affects the learned solution, we differentiate the optimality condition with respect to the weight $w_i$. Differentiating both sides with respect to $w_i$ yields
\begin{equation}
\frac{d}{dw_i}\left[\nabla_\theta \mathcal{L}(\boldsymbol{\theta}^*(\boldsymbol{w}), \boldsymbol{w})\right] = 0
\end{equation}

Expanding this derivative using the chain rule produces two terms: an implicit term describing how the optimal parameters change as the sensor weight varies, and an explicit term describing how the objective function directly depends on the weight. This leads to

\begin{equation}
\label{chainrule}
\underbrace{
\nabla_{\boldsymbol{\theta}}^2
\mathcal{L}\big(\boldsymbol{\theta}^*(\boldsymbol{w}), \boldsymbol{w}\big)
\frac{d\boldsymbol{\theta}^*(\boldsymbol{w})}{d w_i}
}_{\text{implicit term}}
+
\underbrace{
\frac{\partial}{\partial w_i}
\nabla_{\boldsymbol{\theta}}
\mathcal{L}\big(\boldsymbol{\theta}^*, \boldsymbol{w}\big)
}_{\text{explicit term}}
=0
\end{equation}

Recall that the weighted PINN loss is defined in Eq. \ref{weightedloss}. Because the physics loss depends only on the network parameters $\boldsymbol{\theta}$ and does not depend on the sensor weights $\boldsymbol{w}$, the explicit derivative simplifies to

\begin{equation}
\frac{\partial}{\partial w_i}
\nabla_{\boldsymbol{\theta}}
\mathcal{L}(\boldsymbol{\theta}^*,\boldsymbol{w})
=
\nabla_{\boldsymbol{\theta}} l_i(\boldsymbol{\theta}^*).
\end{equation}

It is important to note that the physics constraint still influences the result through the Hessian matrix $\mathbf{H}_{\boldsymbol{\theta}}= \nabla_{\boldsymbol{\theta}}^2
\mathcal{L}\big(\boldsymbol{\theta}^*(\boldsymbol{w}), \boldsymbol{w})$ in the implicit term in Eq. \ref{chainrule}, which contains contributions from both the data loss and the physics loss. Consequently, while the explicit derivative involves only the sensor-specific loss term, the physics constraints still affect the sensitivity of the optimal solution through the curvature of the full objective function. As such, solving for the neural network parameter sensitivity towards $\boldsymbol{w}$ gives

\begin{equation}
\frac{d\boldsymbol{\theta}^*(\boldsymbol{w})}{d w_i}
=
-
\mathbf{H}_{\boldsymbol{\theta}}^{-1}
\nabla_{\boldsymbol{\theta}} l_i\big(\boldsymbol{\theta}^*\big).
\end{equation}

This expression reveals how perturbations of the weight associated with a single sensor propagate through the optimization process and influence the trained neural network parameters. While the above analysis describes how the network parameters change, our ultimate goal is to determine how each sensor affects the accuracy of the reconstructed physical state. Let $E$ denote an error metric defined in the target domain. Based on the forward model defined in Eq.~\ref{forward}, the reconstruction error can be expressed as
\begin{equation}
E(\boldsymbol{\theta}^*(\boldsymbol{w})) =
\left\|
\boldsymbol{y} - \mathcal{F}(\boldsymbol{u}_{\boldsymbol{\theta}})
\right\|^2.
\end{equation}
Here, $\boldsymbol{y}$ and $\boldsymbol{u}_{\boldsymbol{\theta}}$ denote the measurement vector and the predicted internal field over all spatial locations, respectively, where bold symbols indicate vector-valued quantities rather than values at individual nodes. The error metric $E(\boldsymbol{\theta}^*)$ can also be defined in a task-specific manner, as long as it depends on the reconstructed internal field $\boldsymbol{u}_{\boldsymbol{\theta}}$.

The sensitivity of the error metric with respect to the sensor weight $w_i$ at sensor $i$ is obtained via the chain rule, i.e.,
\begin{equation}
\frac{dE}{dw_i}
=
\nabla_{\boldsymbol{\theta}} E(\boldsymbol{\theta}^*)
\frac{d\boldsymbol{\theta}^*}{dw_i}.
\end{equation}

Substituting the previously derived expression for the network parameter sensitivity yields
\begin{equation}
\frac{dE}{dw_i}
=
-
\nabla_{\boldsymbol{\theta}} E(\boldsymbol{\theta}^*)
\mathbf{H}_{\boldsymbol{\theta}}^{-1}
\nabla_{\boldsymbol{\theta}} l_i(\boldsymbol{\theta}^*).
\end{equation}

This quantity measures how strongly the error metric changes when the contribution of sensor $i$ is perturbed. We therefore define the raw \textbf{FOSSA importance score} for sensor $i$ as the magnitude of this sensitivity:
\begin{equation}
S_i =
\left|
\nabla_{\boldsymbol{\theta}} E(\boldsymbol{\theta}^*)
\mathbf{H}_{\boldsymbol{\theta}}^{-1}
\nabla_{\boldsymbol{\theta}} l_i(\boldsymbol{\theta}^*)
\right|.
\end{equation}

Sensors with larger scores exert stronger influence on the reconstruction accuracy and are therefore considered more informative for the inverse problem. By evaluating this score across all candidate sensors, FOSSA provides a theoretically grounded mechanism for identifying observation locations that contribute most effectively to recovering the latent physical state.
This formulation connects sensor selection directly to the optimization dynamics of the physics-informed inverse solver, enabling a principled assessment of sensor importance that goes beyond purely geometric or heuristic placement strategies.

\subsection{FOSSA-Stage2: Confidence Scoring for Reliable Importance Estimation}

Although the optimality-based formulation provides a principled measure of sensor influence, the numerical estimation of this quantity can vary in reliability across observation nodes. In practice, the importance score requires evaluating an inverse-Hessian action that is obtained through iterative linear solves rather than exact matrix inversion. The accuracy of this computation depends on factors such as convergence behavior, residual reduction, and numerical stability of the solver. Due to the ill-conditioned nature of physics-informed inverse problems and the high dimensionality of neural network parameter spaces, some solves may terminate with different levels of numerical accuracy even when the same algorithm and tolerance are used. Consequently, the reliability of the estimated importance score is not uniform across nodes. This motivates the need for an automated mechanism that evaluates the numerical trustworthiness of the sensitivity estimation at each observation location, allowing unreliable estimates to be identified systematically during sensor importance analysis.

The FOSSA importance score is derived from first-order optimality, yet its numerical estimation depends on the quality of the inverse-Hessian solve. For each observation node $i$, the importance computation requires evaluating the inverse-Hessian action $\mathbf{H}_{\boldsymbol{\theta}}^{-1}
\nabla_{\boldsymbol{\theta}} l_i(\boldsymbol{\theta}^*)$, which we approximate using the conjugate gradient (CG) method. Because this solve is performed numerically rather than analytically, the reliability of the resulting importance estimate may vary across nodes, depending on 
residual reduction and gradient consistency.

To quantify the numerical trustworthiness of the importance estimation at each node, we introduce a node-wise confidence score composed of two complementary components: 

\begin{equation}
C_i = C_{S,i} \cdot C_{G,i}, 
\end{equation}
where $C_{S,i}$ measures solver reliability based on CG diagnostics, 
and $C_{G,i}$ evaluates the consistency between the node-wise loss and its gradient magnitude. Specifically,

\subsubsection{Solver reliability $C_{S,i}$}

During the importance estimation step, the inverse-Hessian action is obtained by solving the linear system
\begin{equation}
\mathbf{H}_{\boldsymbol{\theta}} \mathbf{v}_i
=
\nabla_{\boldsymbol{\theta}} l_i(\boldsymbol{\theta}^*),
\end{equation}
where $\mathbf{v}_i$ approximates the inverse-Hessian product required for the sensitivity computation. This system is solved iteratively using the CG method. Let $\mathbf{v}_i^{(k)}$ denote the approximate solution at iteration $k$. The corresponding residual vector is defined as
\begin{equation}
\mathbf{r}_i^{(k)}
=
\nabla_{\boldsymbol{\theta}} l_i(\boldsymbol{\theta}^*)
-
\mathbf{H}_{\boldsymbol{\theta}} \mathbf{v}_i^{(k)} .
\end{equation}

The relative residual of the solve is then defined as
\begin{equation}
r_i^{\mathrm{rel}}
=
\frac{\|\mathbf{r}_i^{(k)}\|}
{\|\nabla_{\boldsymbol{\theta}} l_i(\boldsymbol{\theta}^*)\|}.
\end{equation}

This quantity measures the accuracy of the linear solve. Smaller values indicate that the computed vector $\mathbf{v}_i$ closely satisfies the Hessian equation, while larger values indicate that the approximation error of the inverse-Hessian action remains relatively large.

Let $r_i^{\mathrm{rel}}$ denote the relative residual of the CG solve associated with node $i$. Let $\rho_{\min}$ denote a practical lower bound for achievable residuals and $\rho_{\mathrm{tol}}$ denote the solver stopping tolerance. We first normalize the residual in logarithmic scale
\begin{equation}
t_i
=
\frac{\log r_i^{\mathrm{rel}} - \log \rho_{\min}}
{\log \rho_{\mathrm{tol}} - \log \rho_{\min}}.
\end{equation}

The value is clipped to the interval $[0,1]$ to obtain
\begin{equation}
\tilde{t}_i = \mathrm{clip}(t_i,0,1).
\end{equation}

The solve confidence is then defined as
\begin{equation}
C_{S,i}
=
C_{\min}^{\mathrm{solve}}
+
\big(1-C_{\min}^{\mathrm{solve}}\big)
(1-\tilde{t}_i)^{p_{\mathrm{solve}}}.
\end{equation}
where $C_{\min}^{\mathrm{solve}}$ defines the minimum confidence assigned to solutions that satisfy the convergence criterion. The exponent $p_{\mathrm{solve}}$ controls the shape of the confidence decay. This formulation ensures that nodes achieving very small residuals receive confidence close to one, while nodes whose residual approaches the stopping tolerance $\rho_{\mathrm{tol}} $ retain a minimum confidence level $C_{\min}^{\mathrm{solve}}$. 
The solve confidence $C_{S,i}$ evaluates the numerical accuracy of the CG approximation of the inverse-Hessian action through the final relative residual. Smaller values indicate that the linear system is solved less accurately, implying reduced reliability of the corresponding importance score estimation.







\subsubsection{Gradient-loss consistency confidence $C_{G,i}$}

Let $l_i(\boldsymbol{\theta}^*)$ denote the node-wise loss and $\nabla_{\boldsymbol{\theta}}l_i(\boldsymbol{\theta}^*)$ denote its gradient. We define the mismatch score
\begin{equation}
s_i =
\log \|\nabla_{\boldsymbol{\theta}} l_i(\boldsymbol{\theta}^*)\|
-
\log l_i(\boldsymbol{\theta}^*).
\end{equation}
which represents the local Lipschitz behavior of the loss landscape around that node. Nodes with unusually large values correspond to steep or unstable parameter directions, which are more likely to produce unreliable importance estimates. Let
\begin{equation}
s_{\mathrm{med}} = \mathrm{median}(s),\quad
\mathrm{MAD}_s = \mathrm{median}(|s-s_{\mathrm{med}}|).
\end{equation}

The standardized deviation score becomes
\begin{equation}
z_i^{(g)}
=
\frac{s_i-s_{\mathrm{med}}}{1.4826\,\mathrm{MAD}_s}.
\end{equation}

Only positive deviations are penalized
\begin{equation}
[z_i^{(g)}]_+ = \max(z_i^{(g)},0).
\end{equation}

The gradient confidence is defined as
\begin{equation}
C_{G,i}
=
\exp(-\eta [z_i^{(g)}]_+).
\end{equation}

The gradient-based confidence $C_{G,i}$ evaluates whether the relationship between the node-wise loss magnitude and its gradient magnitude follows the typical scaling behavior observed across sensors. Nodes that deviate substantially from this statistical pattern are treated as outliers, and smaller values of 
$C_{G,i}$ indicate that the corresponding sensitivity estimate may be numerically unstable.

These components provide a reliability-aware confidence score that quantifies the numerical trustworthiness of the importance estimation at each observation node, enabling unstable sensitivity estimates to be automatically identified and mitigated during sensor importance analysis.


\subsection{FOSSA-Stage3: Confidence-Guided Imputation of Unreliable Importance Scores}

Although the confidence score identifies observation nodes whose importance estimates are numerically unreliable, discarding those nodes altogether may introduce discontinuities into the spatial importance map and may also remove useful local structure. To preserve spatial coherence while correcting unreliable estimates, we perform a confidence-guided imputation step on the observation mesh.

Let $S_i$ denote the raw importance score at observation node $i$, and let $C_i \in [0,1]$ denote its associated confidence score. Given a reliability threshold $\tau$, we define the trusted-node set
\begin{equation}
\mathcal{T} = \{ i : C_i \ge \tau \},
\end{equation}
and the unreliable-node set
\begin{equation}
\mathcal{M} = \{ i : C_i < \tau \}.
\end{equation}
Nodes in $\mathcal{T}$ retain their original importance values, while nodes in $\mathcal{M}$ are treated as missing values to be imputed from neighboring observations.

We represent the observation surface as an undirected graph $G=(V,E)$ induced by the outer-surface triangular mesh, where each node corresponds to an observation location and each edge corresponds to mesh adjacency. For each edge $(i,j)\in E$, let $d_{ij}$ denote the geometric edge length between nodes $i$ and $j$, and let $\mathcal{N}(i)$ denote the set of neighbors of node $i$. The imputed importance value at an unreliable node is computed from a weighted average of its neighboring nodes:
\begin{equation}
\tilde{S}_i
=
\frac{\sum_{j\in \mathcal{N}(i)} w_{ij}\,\tilde{S}_j}
{\sum_{j\in \mathcal{N}(i)} w_{ij}},
\qquad i\in\mathcal{M},
\end{equation}
where $w_{ij}$ is an edge-based weighting coefficient. In the present implementation, we adopt inverse-distance weighting
\begin{equation}
w_{ij} = \frac{1}{d_{ij}+\varepsilon},
\end{equation}
where $\varepsilon>0$ is a small constant introduced for numerical stability. To initialize the procedure, all unreliable nodes are assigned the mean importance score over the trusted-node set,
\begin{equation}
\tilde{S}_i^{(0)}
=
\frac{1}{|\mathcal{T}|}\sum_{j\in\mathcal{T}} S_j,
\qquad i\in\mathcal{M},
\end{equation}
while the trusted nodes remain fixed,
$
\tilde{S}_i = S_i, i\in\mathcal{T}
$.
We then iteratively update only the unreliable nodes using neighboring values on the graph. At iteration $k+1$, the update rule is
\begin{equation}
\tilde{S}_i^{(k+1)}
=
\frac{\sum_{j\in \mathcal{N}(i)} w_{ij}\,\tilde{S}_j^{(k)}}
{\sum_{j\in \mathcal{N}(i)} w_{ij}},
\qquad i\in\mathcal{M}.
\end{equation}
The iteration continues until convergence, defined by
\begin{equation}
\max_{i\in\mathcal{M}}
\left|
\tilde{S}_i^{(k+1)}-\tilde{S}_i^{(k)}
\right|
< \delta,
\end{equation}
where $\delta>0$ is a prescribed stopping tolerance.

This procedure can be interpreted as a graph-based harmonic extension of the trusted importance values into low-confidence regions. In this way, the final imputed importance map preserves the reliable structure identified by the confidence score while replacing unstable local estimates with spatially consistent values inferred from neighboring nodes on the observation mesh.

\section{Physics-Constrained Inverse ECG Framework and Experimental Setup}

\subsection{Cardiac Electrophysiological Model}

To characterize the underlying cardiac electrophysiology (EP) rule, we adopt the Aliev--Panfilov (AP) model, which has been widely used to describe the spatiotemporal evolution of cardiac electrophysiology. Let $u(\boldsymbol{x}_h,t)$ denote the normalized heart surface potential (HSP), and $v(\boldsymbol{x}_h,t)$ represent the recovery variable governing local depolarization behavior, where $\boldsymbol{x}_h$ and $t$ represent the spatial coordinate of the heart and temporal instance, respectively. The AP model is formulated in partial differential equations (PDEs):
\begin{align}
\label{APmodel}
\frac{\partial u}{\partial t}
&= \nabla \cdot (D \nabla u) + k_r u(u-a)(1-u) - uv, \nonumber\\
\frac{\partial v}{\partial t}
&= \xi(u,v)\left(-v - k_r u(u-a-1)\right), \nonumber\\
\mathbf{n} \cdot \nabla u \big|_{\partial \mathcal{H}} &= 0,
\end{align}
where $\nabla$ denotes the spatial gradient operator defined on the heart geometry, $D$ is the diffusion conductivity, $k_r$ controls the repolarization rate, and $a$ determines tissue excitability. The nonlinear coupling term is defined as $\xi(u,v) = e_0 + \mu_1 v / (u + \mu_2)$. The Neumann boundary condition ensures no current flows out of the heart surface $\partial \mathcal{H}$. This electrophysiological model provides a physics-based description of cardiac wave propagation and will be used to constrain the inverse ECG reconstruction.

\subsection{Inverse ECG Problem Formulation}

The inverse ECG problem aims to reconstruct the heart surface potential $u(\boldsymbol{x},t)$ from body surface potential measurements $y(\boldsymbol{x},t)$. The forward relationship between the heart and body surface is governed by a transfer matrix $\boldsymbol{R}$, given by:
\begin{equation}
y(\boldsymbol{x}_b,t) = \boldsymbol{R} u(\boldsymbol{x}_h,t),
\end{equation}

where $\boldsymbol{x}_b$ denotes the body spatial coordinate and $\boldsymbol{R} \in \mathbb{R}^{N_b \times N_h}$ is the transfer matrix between the heart and torso, which can be derived from Divergence Theorem and Green’s Second Identity \cite{yao2020spatiotemporal, barr2007relating}. Due to the ill-conditioned nature of $\boldsymbol{R}$, directly inverting this relationship leads to unstable and non-unique solutions. Therefore, additional regularization is required to achieve robust reconstruction. The error metric adopted in the current investigation is the body--heart transformation error, formulated as
\begin{equation}
E = \| y(\boldsymbol{x}_b,t)  - \boldsymbol{R} \hat{u}(\boldsymbol{x}_h,t;\boldsymbol{\theta}^*) \|^2,
\end{equation}
which quantifies the mismatch between measured body-surface signals and those reconstructed via the forward operator from the estimated cardiac field at the converged network parameters $\boldsymbol{\theta}^*$.

\subsection{Physics-Aware PINN Formulation}

To address the ill-posedness of the inverse ECG problem, we adopt a PINN framework. Specifically, we parameterize the unknown spatiotemporal mapping using a feed forward neural network ${\mathcal{N}(\boldsymbol{x}_h,t;\theta_{NN})}$. The network is trained by minimizing a composite loss function:
\begin{equation}
\mathcal{L}(\theta_{NN}) = \lambda_d\mathcal{L}_{hb} + \lambda_p \cdot \mathcal{L}_{ph},
\end{equation}
where $\mathcal{L}_{hb}$ is the data-driven loss enforcing consistency with body surface measurements, and $\mathcal{L}_{ph}$ is the physics-based loss derived from the AP model. The data-driven loss is defined as:
\begin{equation}
\mathcal{L}_{hb}
=  \sum_{\boldsymbol{x}_b,t} \left\| y(\boldsymbol{x}_b,t) - \boldsymbol{R} \hat{u}(\boldsymbol{x}_h,t) \right\|^2,
\end{equation}
which enforces agreement between predicted heart potentials and observed body surface potentials. The physics-based loss incorporates the residuals of the AP model:
\begin{equation}
\mathcal{L}_{f}
= \frac{1}{N_f} \sum_{i=1}^{N_f}
\left( r_u(x_i,t_i)^2 + r_v(x_i,t_i)^2 + r_b(x_i,t_i)^2 \right)
\end{equation}
where $r_u$, $r_v$, and $r_b$ denote the PDE residuals corresponding to the governing equations in Eq. \ref{APmodel}. $N_f$ is the number of collocation points that are selected from the cardiac spatiotemporal domain. 
Automatic differentiation is employed to compute the required temporal and spatial derivatives with respect to $[\boldsymbol{x}_h,t]$, enabling exact enforcement of the governing PDE constraints.

\begin{figure*}[t]
\centering
\includegraphics[width=\textwidth]{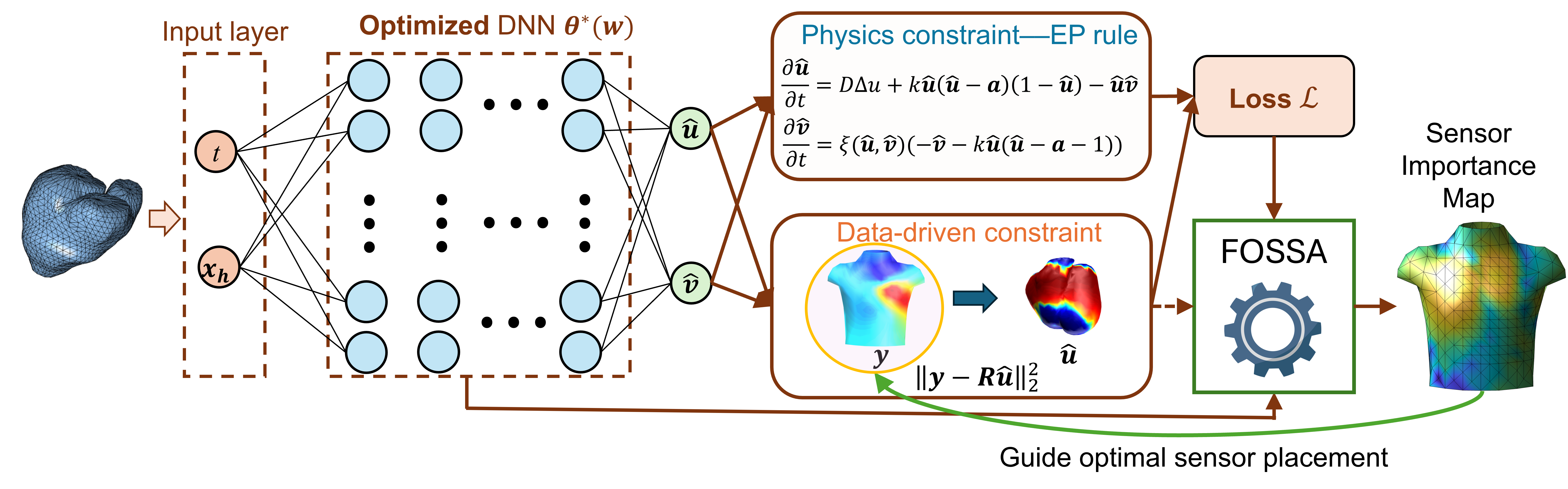}
\caption{Illustration of the proposed FOSSA-integrated physics-constrained inverse ECG framework. }
\label{fig:fossa_pipeline}
\end{figure*}

While the above physics-aware PINN framework enables robust reconstruction of the heart-surface potential from body-surface measurements, its performance critically depends on the spatial distribution of the sensing locations. In practical settings, the number of available ECG sensors is inherently limited, which necessitates a customized distribution of sensing locations to maximize reconstruction accuracy. Rather than uniformly or heuristically placing sensors, the objective is to identify an optimal subset of body-surface nodes that can most effectively capture the underlying cardiac dynamics for a given heart system. To address this, we propose the FOSSA strategy, whose goal is to identify the most informative measurement nodes for a given cardiac system. Specifically, as illustrated in Fig.~\ref{fig:fossa_pipeline}, FOSSA performs post-training sensitivity analysis on the converged PINN model to quantify the contribution of each candidate sensing location to the inverse reconstruction and selects a subset of nodes that maximizes the informativeness of the measurements. This enables an efficient and principled sensor placement scheme tailored to the individual electrophysiological dynamics.

\begin{figure*}[t]
\centering
\includegraphics[width=5in]{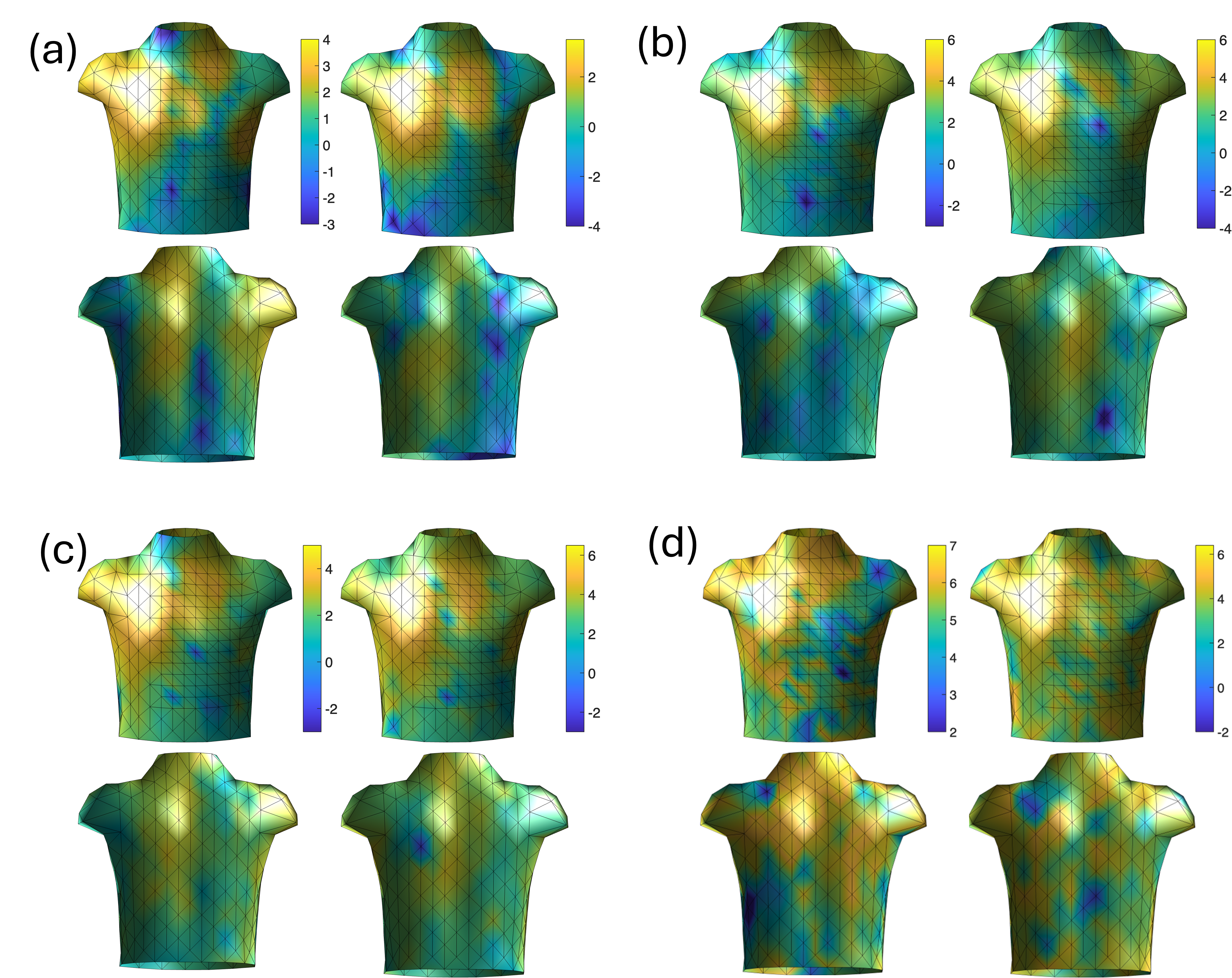}
\caption{Reproducibility of FOSSA-derived importance maps under different noise levels and training conditions. Each subfigure corresponds to a noise setting: (a) no noise, (b) $\sigma=0.005$, (c) $\sigma=0.01$, and (d) $\sigma=0.05$. Front and back views of the body surface are shown. }
\label{fig:fossa_reproducibility}
\end{figure*}

\begin{figure*}[t]
\centering
\includegraphics[width=\textwidth]{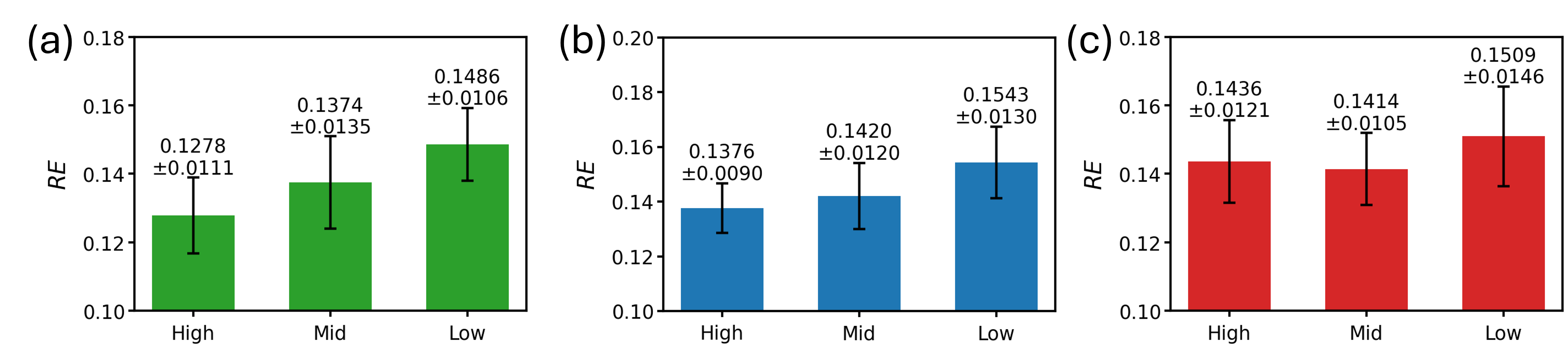}
\caption{Reconstruction performance under different FOSSA-ranked observation sets and noise levels. (a) $\sigma=0.005$, (b) $\sigma=0.01$, and (c) $\sigma=0.05$. For each noise level, three observation sets are constructed based on FOSSA importance ranking: high-, middle-, and low-importance nodes.}
\label{fig:fossa_selection}
\end{figure*}

\begin{figure}[t]
\centering
\includegraphics[width=3.5in]{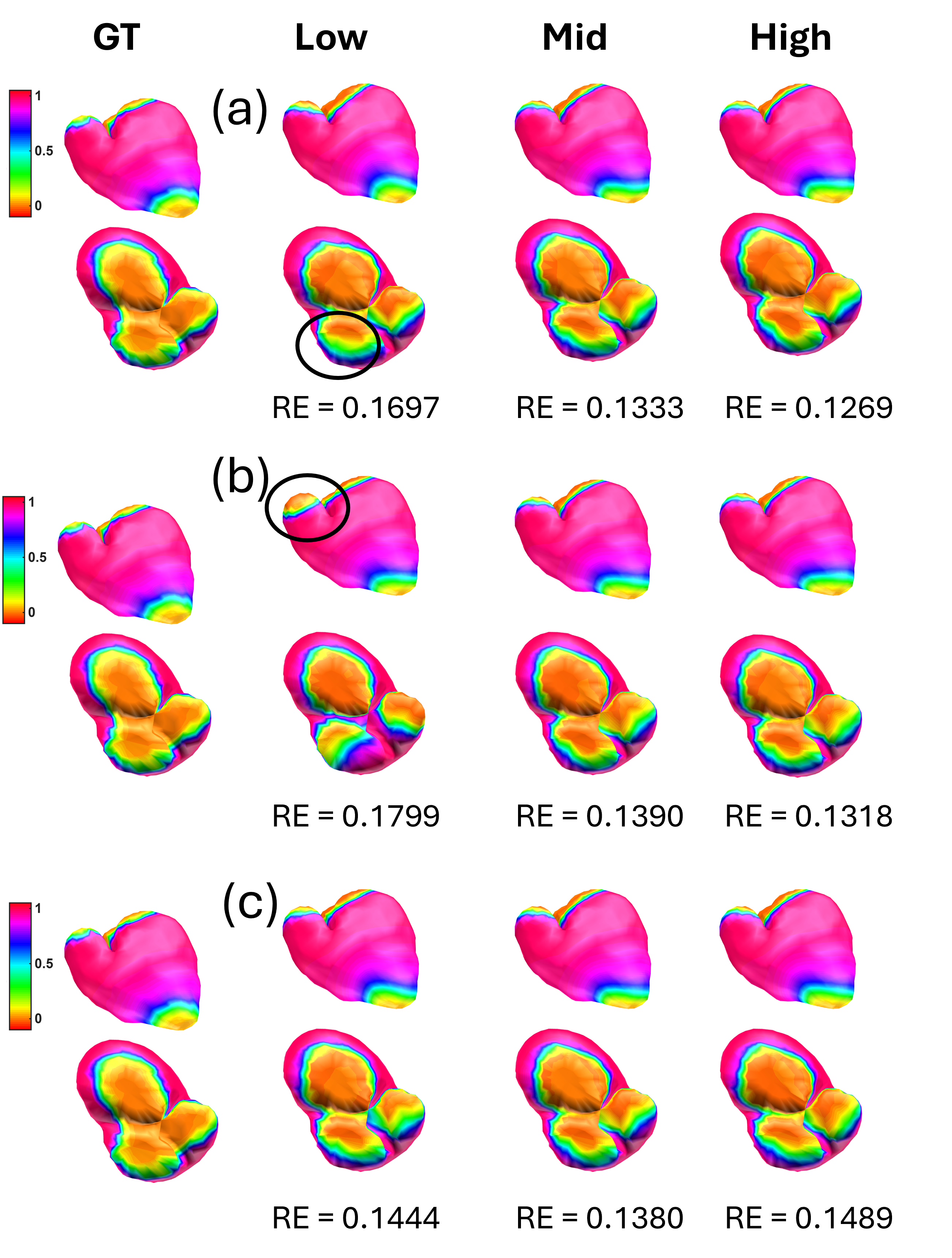}
\caption{Qualitative comparison of heart-surface potential reconstruction at a representative temporal instance under different FOSSA-ranked observation sets and noise levels at different noise level: (a) $\sigma=0.005$, (b) $\sigma=0.01$, and (c) $\sigma=0.05$. }
\label{fig:fossa_qualitative}
\end{figure}

\section{Experimental results}
\label{Section: results}

\subsection{Reproducibility of FOSSA Importance Maps Under Noise and Training Variability}
To evaluate the reproducibility and robustness of the proposed FOSSA strategy, we investigate the consistency of the derived importance maps under varying noise levels and training stochasticity. Specifically, we consider four noise settings added to the body-surface potential measurements: (a) no noise, (b) Gaussian noise with standard deviation $\sigma = 0.005$, (c) $\sigma = 0.01$, and (d) $\sigma = 0.05$. For each noise level, multiple PINN models are trained using different random initializations and distinct sets of collocation points for enforcing the electrophysiological constraints.

Fig.~\ref{fig:fossa_reproducibility} presents the resulting importance maps generated by FOSSA on both the front and back views of the body surface. Across low to moderate noise levels (i.e., $\sigma = 0$, $0.005$, and $0.01$), the importance maps exhibit consistent spatial patterns, with similar distributions observed across different training instances. In particular, regions with high importance scores remain stable, with dominant contributions concentrated in specific anatomical areas, most notably along the upper-left torso region (from the human anatomical perspective), while low-importance regions consistently appear in less informative zones. 

However, when the noise level increases to $\sigma = 0.05$, the estimated importance maps become noticeably less stable, showing increased variability and less coherent spatial structure. The underlying reason is that the importance scores are derived from sensitivity with respect to the observed data, and high noise levels can perturb the learned solution of the inverse problem, leading to amplified fluctuations in the computed gradients. This degradation indicates that while the FOSSA-derived sensitivity is robust to low and moderate noise and is not dominated by training randomness, its reliability diminishes under high-noise conditions. Nevertheless, the consistency observed across different random initializations and collocation sampling at lower noise levels suggests that the proposed method captures intrinsic properties of the inverse ECG system rather than artifacts of the optimization process.



These results confirm that FOSSA provides a reproducible and robust estimation of sensor importance, making it suitable for guiding reliable sensor placement under practical conditions where measurement noise and training variability are unavoidable.

\subsection{Effect of FOSSA-Ranked Body Observations on Inverse Cardiac Reconstruction}

To further evaluate whether the FOSSA-derived importance scores are meaningful for practical sensor selection, we design an experiment to examine how different rank-based body-surface observation sets affect the reconstruction of HSP. The central hypothesis is that body sensors assigned higher importance by FOSSA should provide more informative observations for the inverse ECG problem, thereby leading to more accurate PINN-based reconstruction.

In this study, the body surface consists of 352 candidate sensing nodes. After obtaining the FOSSA importance score for each node, all body observations are ranked in descending order according to their estimated importance. Based on this ranking, we construct three observation sets with the same sensing budget $k=252$: a high-importance set formed from the top-ranked body nodes, a middle-importance set formed from nodes with intermediate ranks, and a low-importance set formed from the lowest-ranked body nodes. These three sets are then used separately as the available body-surface observations for solving the inverse ECG problem.

The reconstruction performance is evaluated by comparing the predicted HSP with the ground-truth cardiac dynamics. Specifically, we use the relative error (RE) as the primary metric, which quantifies the discrepancy between the reconstructed and reference heart-surface potential over the full spatiotemporal domain:
The relative error is defined as
\begin{equation}
RE =
\frac{\sqrt{\sum_{\boldsymbol{x}_h,t}\left(\hat{u}(\boldsymbol{x}_h,t)-u(\boldsymbol{x}_h,t)\right)^2}}
{\sqrt{\sum_{\boldsymbol{x}_h,t}u(\boldsymbol{x}_h,t)^2}},
\end{equation}
where $u(\boldsymbol{x}_h,t)$ and $\hat{u}(\boldsymbol{x}_h,t)$ denote the ground-truth and reconstructed heart-surface potential, respectively. Since body-surface measurements are inevitably contaminated by noise in practical settings, we do not consider the noise-free case in this experiment. Instead, the evaluation is conducted under three noise levels, namely $\sigma = 0.005$, $0.01$, and $0.05$, in order to assess how the effectiveness of FOSSA-based sensor ranking changes with observation quality.


The reconstruction performance corresponding to different FOSSA-ranked observation sets is summarized in Fig.~\ref{fig:fossa_selection} for three noise levels. A clear and consistent trend can be observed: the high-importance observation set achieves the lowest reconstruction error across all noise conditions, while the low-importance set yields the highest error. The middle-ranked set consistently falls between these two extremes.

At lower noise levels ($\sigma = 0.005$ and $0.01$), the performance gap between the high- and low-importance sets is more pronounced. For example, in Fig.~\ref{fig:fossa_selection}(a), the high-importance set achieves a relative error of $0.1278 \pm 0.0111$, compared to $0.1486 \pm 0.0106$ for the low-importance set. A similar trend is observed in Fig.~\ref{fig:fossa_selection}(b), where the reconstruction error increases from $0.1376 \pm 0.0090$ (high) to $0.1543 \pm 0.0130$ (low). These results demonstrate that selecting sensors based on FOSSA-derived importance scores leads to substantially improved reconstruction accuracy under low to moderate noise conditions.

As the noise level increases to $\sigma = 0.05$, the previously observed monotonic trend begins to break down. As shown in Fig.~\ref{fig:fossa_selection}(c), the performance gap between the high-, middle-, and low-importance observation sets becomes less distinct, and the relative errors are no longer strictly ordered according to the importance ranking. In addition, the variance increases, indicating reduced stability in reconstruction performance.
This behavior is consistent with the earlier observation that high noise levels degrade the stability of the importance estimation. Consequently, the advantage of importance-guided sensor selection is reduced when the measurements are heavily corrupted by noise.

Overall, these results confirm that the FOSSA-derived importance scores are strongly correlated with the informativeness of body-surface observations. Sensors with higher importance scores contribute more effectively to the inverse reconstruction, validating the effectiveness of FOSSA as a principled strategy for sensor selection under realistic noise conditions.

Fig.~\ref{fig:fossa_qualitative} presents qualitative comparisons of the reconstructed heart-surface potential at a representative temporal instance under different FOSSA-ranked observation settings and noise levels. The ground truth is shown alongside reconstructions prediction obtained using high-, middle-, and low-importance body-surface observation sets.

Overall, the reconstruction obtained from the high-importance observation set closely matches the ground-truth spatial patterns, accurately capturing both the global structure and local features of the cardiac potential distribution. In contrast, reconstructions based on lower-ranked observation sets exhibit noticeable deviations, particularly in regions with sharp spatial transitions. These discrepancies are highlighted in the figure (e.g., circled regions), where low-importance observations fail to recover fine-scale structures and exhibit spatial smoothing or distortion.
As the noise level increases, the reconstruction quality degrades across all observation sets, and the visual distinction between different rankings becomes less pronounced. This observation is consistent with the quantitative results, where the advantage of importance-based sensor selection diminishes under high-noise conditions.
These qualitative results further validate that FOSSA-derived importance scores effectively identify informative sensing locations that are critical for accurately reconstructing heart-surface dynamics, particularly in capturing localized electrophysiological features.

\begin{figure*}[t]
\centering
\includegraphics[width=5in]{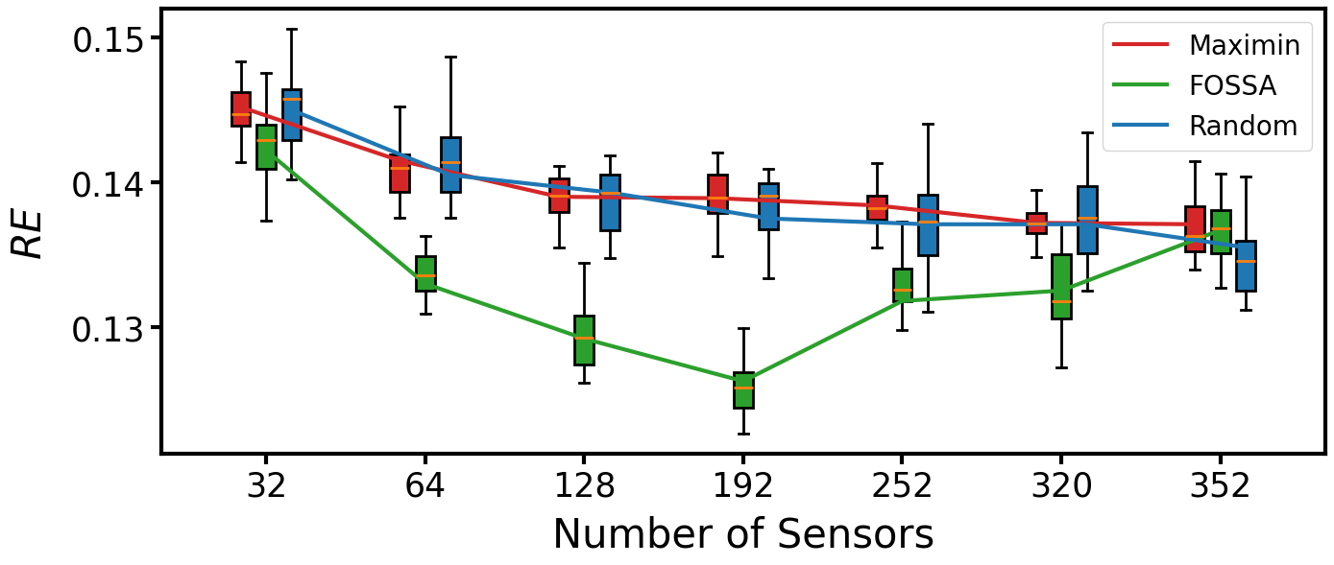}
\caption{Comparison of sensor selection strategies for inverse ECG reconstruction under varying sensing budgets ($\sigma = 0.01$).}
\label{fig:selection_strategies}
\end{figure*}

\subsection{Comparison of Sensor Selection Strategies for Inverse Reconstruction}

To evaluate the effectiveness of different body-surface sensor selection strategies for inverse ECG reconstruction, we compare three representative approaches in terms of their impact on HSP reconstruction accuracy. In this experiment, the observation noise is fixed at $\sigma = 0.01$, representing a moderate noise level.
We consider three sensor selection strategies: 
\begin{itemize}
    \item \textit{Random selection}, where a subset of body-surface sensors is randomly sampled without any prior knowledge. 
    \item  \textit{Maximin distance-based strategy}, which aims to maximize spatial coverage: starting from an initial sensor, additional sensors are iteratively selected such that each newly chosen location maximizes the minimum geodesic distance to the existing set of selected sensors. This results in a set of sensors that are as spatially dispersed as possible over the body surface \cite{xie2022physics1, johnson1990minimax}. 
    \item \textit{FOSSA-based selection}, sensors are ranked according to their importance scores derived from the proposed FOSSA, and the top-ranked sensors are selected sequentially.
\end{itemize}

For each selection strategy, we vary the number of deployed sensors from 32 to 352, allowing a comprehensive evaluation across different sensing budgets. For each configuration, the selected body-surface observations are used to solve the inverse ECG problem via the PINN framework, and the reconstruction accuracy is evaluated using the RE between the predicted and ground-truth HSP.

Fig. \ref{fig:selection_strategies} compares the reconstruction performance of different sensor selection strategies under varying sensing budgets. Overall, both the random and maximin strategies exhibit relatively similar trends: as the number of sensors increases, the reconstruction error gradually decreases and then stabilizes, indicating improved coverage of the body surface.

In contrast, the FOSSA-based selection demonstrates a distinct non-monotonic behavior. At lower sensor budgets, FOSSA consistently achieves significantly lower reconstruction error compared to both random and maximin strategies. This indicates that selecting sensors based on importance ranking effectively prioritizes the most informative locations, leading to more accurate inverse reconstruction with limited observations. As the number of sensors increases, the reconstruction error of FOSSA decreases first and reaches its minimum at sensing budget of 192. However, beyond this point, the performance begins to degrade slightly as more sensors are included. This behavior suggests that while high-importance sensors provide strong constraints for the inverse problem, incorporating lower-ranked sensors may introduce less informative or redundant observations, which can negatively affect the reconstruction under noisy conditions.

Eventually, as the number of sensors approaches the full set, the performance of all strategies converges, since the observation becomes nearly complete and the effect of selection diminishes.
Overall, these results demonstrate that FOSSA is particularly advantageous in the low- to moderate-sensor regime, where careful selection of informative sensing locations is critical. The observed performance trend also highlights that importance-based selection is most effective when focusing on the most informative subset, rather than simply increasing the number of sensors.

\section{Conclusions}
\label{Section: conclusions}
In this work, we introduced FOSSA, a first-order optimality-based sensitivity-driven sensor selection framework for inverse PINNs, and demonstrated its effectiveness in electrocardiogram imaging. By leveraging first-order sensitivity of the pre-trained PINN objective with respect to observation locations, FOSSA provides a principled mechanism to quantify the contribution of each sensing node to the reconstruction. Through a series of experiments, we showed that the resulting importance maps are reproducible under low to moderate noise and capture stable, anatomically meaningful patterns. More importantly, sensor selection based on FOSSA ranking leads to consistently improved reconstruction accuracy compared to random and geometry-based strategies, particularly in the low- to moderate-sensor regime where observation resources are constrained. The results also reveal a non-monotonic behavior, indicating that incorporating lower-ranked sensors may introduce redundancy or noise that degrades performance, highlighting the importance of selective measurement rather than dense sampling. While the advantage of FOSSA diminishes under high noise levels due to instability in sensitivity estimation, the overall findings demonstrate that informative measurements, rather than uniform coverage, play a dominant role in inverse reconstruction. This work establishes a direct link between sensitivity analysis and sensor placement in PINN-based inverse problems and provides a foundation for developing adaptive and system-aware sensing strategies in more complex settings.

\bibliographystyle{IEEEtran}
\bibliography{ref}

\vfill

\end{document}